\documentclass[10pt,english]{revtex4}
\usepackage{lmodern}
\usepackage{graphicx}

\usepackage[T1]{fontenc}
\usepackage[latin9]{inputenc}
\usepackage[a4paper]{geometry}
\usepackage{color}
\usepackage{babel}
\usepackage{float}
\usepackage{units}
\usepackage{url}
\usepackage{amsthm}
\usepackage{amsmath}
\usepackage{amssymb}
\usepackage{natbib}

\begin{document}

\title{Nonsingular cosmology from evolutionary quantum gravity.}
\author{Francesco Cianfrani}
\email{francesco.cianfrani@ift.uni.wroc.pl}
\affiliation{Institute
for Theoretical Physics, University of Wroc\l{}aw, Plac\ Maksa Borna
9, Pl--50-204 Wroc\l{}aw, Poland.}
\author{Giovanni Montani}
\email{giovanni.montani@frascati.enea.it}
\affiliation{ENEA, Unit\`a Tecnica Fusione, ENEA Centro Ricerche Frascati, via Enrico Fermi 45, 00044 Frascati (Roma), Italy}
\author{Fabrizio Pittorino}
\email{fabrizio.pittorino@gmail.com}
\affiliation{Universit\`a di Roma ``Sapienza'', Piazzale Aldo Moro 5, 00185 Roma, Italy}

\begin{abstract}
We provide a cosmological
implementation of the evolutionary
quantum gravity, describing an isotropic Universe,
in the presence of a negative
cosmological constant and a
massive (preinflationary)
scalar field.
We demonstrate that the considered
Universe has a nonsingular
quantum behavior, associated to
a primordial bounce, whose ground state
has a high occupation number.
Furthermore, in such a vacuum state,
the super-Hamiltonian eigenvalue
is negative, corresponding to
a positive emerging
dust energy density.
The regularization of the model is
performed via a polymer quantum
approach to the Universe scale factor
and the proper classical limit is
then recovered, in agreement with
a preinflationary state of the Universe.
Since the dust energy density is
redshifted by the Universe deSitter
phase and the cosmological constant
does not enter the ground state
eigenvalue, we get a late-time
cosmology, compatible with
the present observations, endowed with
a turning point in the far future.
\end{abstract}

\maketitle

\section{Introduction}

The problem of defining a proper
notion of time in quantum gravity
is one of its most puzzling
questions, common to the different
existing approaches \cite{Rovelli:1990ph,Isham:1992ms,libro}. From a physical point of view,
the relational approach \cite{Rovelli:1990pi}
appears very promising, especially
when matter is included in
quantum dynamics. In fact, as discussed in
\cite{Kuchar:1990vy,Brown:1994py,Mercuri:2003wn} a dualism
exists between a dust fluid
and the time evolution of the
quantum gravitational field.

However, such a promising approach
has a weak point, concerning the
nonpositive character of the
super-Hamiltonian spectrum in
comparison with the intrinsic
positive energy density associated
to a dust fluid.

The possibility to interpret the
super-Hamiltonian eigenvalues
as the comoving contribution
$-\rho \sqrt{h}$
($\rho$ being the dust energy density
and $h$ the three-metric determinant)
allows interesting speculations
\cite{Savchenko:1998ru,Montani:2002ca,Montani:2008xc} on the role played in
quantum gravity by a reference frame:
no longer a simple gauge reparametrization
of the dynamics, but a real source
involved in the system evolution,
{\it i.e.} a quantum violation of the
general relativity principle \cite{Montani:2009wg}. Such a dualism between a dust fluid
and a clock has a mandatory implementation
in quantum cosmology, where the
nonvanishing super-Hamiltonian
spectrum must provide, on the
classical limit, a dustlike
cosmological component of the Universe,
also investigated as a possible
dark matter candidate \cite{Corvino:2004wt,Battisti:2006uj}.

Here we provide a self-consistent
picture of the quantum evolution of
the early Universe, as described in the
framework of evolutionary quantum
gravity, showing how the resulting
dust component has a positive energy
density and its contribution to the
Universe critical parameter is redshifted
by the inflationary scenario to unobservable values. We consider a homogeneous and isotropic
flat Universe, endowed with a negative
cosmological constant and a massive
scalar field, well mimicking the
preinflationary behavior of the inflaton field \cite{KolbTurner,Baumann:2009ds,Montani:2011zz} in a model with a decay from a false vacuum (see \cite{Bousso:2013uia,Bousso:2014jca} for the comparison of this scenario with experimental data).

We analyze the quantum evolution of this
cosmological model according to a revised
Wheeler-DeWitt approach which allows
for a time evolution of the Universe
wave function, as determined by a Schr\"odinger
prescription.
By an adiabatic approximation, we
show how the considered quantum dynamics
has a well-defined classical limit
and at early stages predicts
a nonsingular behavior
(removing the initial singularity),
in close analogy to a big-bounce
structure in the limit of very small
Universe volumes.
Furthermore, we calculate the full
spectrum of the super-Hamiltonian,
which is associated to a positive 
value of the dual dust energy density,
but with the shortcoming of an unbounded-from-below profile of the eigenvalues.

To remove the unpleasant feature of
an unstable quantum system which
does not possess a ground state,
we treat the Universe volume as
a discrete variable in
the polymer quantum approach \cite{Ashtekar:2002sn,Corichi:2007tf,Battisti:2008bv}. 
As a result, the positive nature of
the spectrum is altered
(states corresponding to negative dust energy density appear),
but a stable ground state emerges
in correspondence to very high
occupation numbers.
Such a ground (vacuum) state is
associated to a positive dust energy density
and it is suitable to implement
a quasiclassical limit of the dynamics.
In fact, in the limit $\hbar \rightarrow 0$,
the high occupation number of this ground state ensures
finite energy density, which is 
of Planckian value if the discretization
parameter of the polymer approach is
taken of the Planck length order.
It is worth stressing how the value
of the negative cosmological constant
does not enter the ground state
eigenvalue and it can be taken
sufficiently small to ensure a
turning point of the Universe in the far
future (de facto the presented model
describes a cyclic Universe,
possessing a future classical turning
point and a bounce in the past).

In the proposed scheme, the
preinflationary Universe emerges
as the classical limit of an evolutionary
quantum dynamics and it is endowed
by a dust energy density (relic
of the quantum clock) which is
then redshifted to very small
values by the subsequent de Sitter phase.
We stress how the dust contribution,
differently from the ordinary matter,
cannot be restored by the inflaton decay 
during the reheating phase and
its fate is an increasing dilution
up to present unobservable amounts.
In other words, the impossibility
to observe today the matter counterpart
of the evolutionary quantum gravity
model is explained in the same
spirit as the ``unwanted relic paradox''
is solved by inflation \cite{KolbTurner,Montani:2011zz}.

Therefore, we provide a very promising
cosmological implementation of the
evolutionary quantum gravity,
thought as the intrinsic matter-time
dualism, which relies on the
existence of a regular ground state
of the Universe, endowed with
appropriate properties for a regular
classical limit.

In summary, while the problem of introducing a 
reliable time variable in quantum 
gravity is clearly an open question, 
and many competitive approaches stand 
in literature \cite{Isham:1992ms,libro,Rovelli:2004tv,Thiemann:2007zz}, 
we emphasize how the present analysis 
enforces the idea that the time-fluid 
dualism admits intriguing cosmological 
applications, as already inferred in 
\cite{Battisti:2006uj}. 
In fact, the most significant difficulty 
of such a relational approach is the nonpositive nature of the
super-Hamiltonian spectrum, resulting in the nonpositive 
energy density of the fluid.
We propose as solution the restriction 
of such a requirement to the ground state 
of the theory only.
Here, we achieve a positive regularized 
``vacuum'' state for the model, via 
the introduction of a negative 
cosmological constant, but its 
validity can be more general.
In this sense, the matter-time dualism 
possesses a nice feature in the 
cosmological paradigm, which 
upgrades its reliability, making it 
of comparable impact to the well-known 
multitime approach \cite{Isham:1984rz}, 
which is very powerful for characterizing the 
cosmological evolution in terms 
of the Universe volume as time variable.

The present paper is organized as follows: in Sec. \ref{II} we present the framework of evolutionary quantum gravity in a minisuperspace context; in Sec. \ref{IIbis} we give some physical arguments on why we consider an evolutionary quantum cosmology and a negative cosmological constant; in Sec. \ref{III} we introduce the cosmological model and we solve the associated evolutionary Schr\"odinger equation in the early- and late-time limits; in Sec. \ref{IV} we perform polymer quantization as $\rho\rightarrow0$ and we outline the emergence of a bounded-from-below energy spectrum. Finally, in Sec. \ref{V} brief conclusions follow. 

\section{Evolutionary quantum cosmology}\label{II}

One of the most promising approaches
to the problem of time in quantum
gravity relies on the dualism existing
between a dust fluid and a physical
clock \cite{Kuchar:1990vy,Brown:1994py,Mercuri:2003wn}. Here
we briefly discuss this correspondence
in the framework of the minisuperspace,
{\it i.e.} as restricted to the case of
homogeneous cosmological models.

The cosmological implementation
of the Wheeler-DeWitt equation corresponds
to dealing with a finite number of
degrees of freedom, say $q_i$
($i=1,2,...,n$) generalized coordinates,
representing scale factors of the
Universe and matter fields.
The dynamics of the model is
summarized by the classical system
Hamiltonian $H(q_i,p_i)$,
$p_i$ being the conjugate momenta
to the generalized coordinates.
Implementing an evolutionary quantum
dynamics for the considered homogeneous
model, in place of the standard Wheeler-DeWitt
frozen formalism \cite{Kuchar:1980ht},
consists of assuming that the
Universe wave function $\psi$ evolves with
respect to an external parameter $t$,
which plays the role of a physical
clock; {\it i.e.} we take $\psi = \psi (t,q_i)$.

The evolution of the system is then naturally 
determined by the Schr\"odinger equation
\begin{equation}
i\hbar \partial _t \psi = N(t)\hat{\mathcal{H}} \psi
\, ,
\label{sceq}
\end{equation}
where $N(t)$ denotes the lapse
function and $\hat{\mathcal{H}}$ is the operator version of the super-Hamiltonian.
By taking the wave function in the
following integral representation
\begin{equation}
\psi (t, q_i) = \int dE\,
\phi (E , q_i)
\exp \left\{ -\frac{i}{\hbar} E\int N(t)dt\right\}
\, ,
\label{inre}
\end{equation}
the Schr\"odinger equation above 
is associated to the time independent
eigenvalue problem
\begin{equation}
\hat{\mathcal{H}}\phi = E \phi,
\end{equation}
$E$ being the super-Hamiltonian
eigenvalue.

As far as we take the classical limit,
for $\hbar\rightarrow 0$, by setting the
wave function as
$\psi \sim \exp{i\sigma/\hbar}$, 
the eigenvalue problem above takes
the form of an Hamilton-Jacobi equation,
containing an additional matter contribution
$-E$, $\sigma$ being the Jacobi function.
Since the Hamiltonian is a scalar
density of weight $1/2$, to get an
energy density, we need to divide this
new contribution by the Universe volume
$V_u\sim \sqrt{h}$ ($h$ being the
three-metric determinant) and hence
we get $\rho _{new} =        
-E /V_u$, which clearly describes
a dust comoving fluid.

The clock-dust-fluid dualism is
well expressed in quantum cosmology
by the correspondence traced above
between the super-Hamiltonian spectrum
and the energy density of the dust emerging in
the classical limit of the evolutionary
quantum picture.
The limit of this analogy is in the
nonpositive nature of the super-Hamiltonian
spectrum, which prevents us from ensuring an
always positive dust energy density.
Indeed, we stress how, invoking a
minimal energy principle for the quantum
Universe, the request of a positive dust fluid
energy density must be transferred
to such a ground (vacuum) state only.
In this respect, we finally observe how
the value of the emerging contribution $\rho_{new}$
depends directly on the boundary conditions,
characterizing the considered cosmological
model, which fixes the ground state
eigenvalue. As a result, it is
not immediately recognizable that the constraint
ensuring that the emerging dust behaves
as a test fluid. In this sense, the
emerging energy density must be regarded as
the physical substantiation of the
comoving reference frame.
In other words, an evolutionary
quantum cosmology predicts, in the
classical limit, the existence of a natural
preferred comoving reference frame,
unavoidably affecting the quantum evolution
of the system. Despite the fact that the coordinate
reparametrization of the adopted scheme is
still allowed, the present scenario
can be interpreted as quantum breaking of
the general relativity principle \cite{Montani:2008xc}.

\section{Physical grounds and motivation}\label{IIbis}

We now address two subtle features 
at the ground of our model, which are 
indeed intrinsically connected to 
each other, {\it i.e.} the implementation 
of an evolutionary quantum cosmology 
and the presence of a negative 
cosmological constant in the Universe 
dynamics. 

Clearly, the solution of the problem of time 
is one of the open questions in quantum gravity, 
both for what concerns the nature of the field, 
materializing the clock, and also for the properties 
time must possess on a classical regime, as well as 
on a quantum regime. This problem was interestingly 
addressed in \cite{Rovelli:1990ph,Rovelli:1989jn,Rovelli:1990pi}, where 
the relation between defining a time and defining a 
reference frame in quantum gravity are related concepts. 
For some attempts to characterize the time variable 
as a relational clock having peculiar properties, like a monotonic behavior, 
see \cite{Thiemann:2006up,Montani:2008jm}. 
However, one of the most interesting points of view 
on this topic, different from the one here proposed, 
is the so-called multitime approach \cite{Isham:1984rz}, 
in which it is emphasized how the gravitational field has to 
be separated into its two real physical degrees of freedom, 
while the remaining part of the space geometry labels 
the evolution. For a review of the various 
viable approaches to the problem of time in quantum 
gravity and a discussion of their successes and 
shortcomings, see \cite{Isham:1992ms} 
(see also \cite{libro}).

However, as discussed in the previous section, 
the dualism between an 
external time and the presence of 
a dust fluid in the quantum dynamics is 
a well-established fact. The physical 
nature of time depends  
on the details of the considered model 
and it relies on the link 
existing between the super-Hamiltonian 
eigenvalue and the dust energy density 
\cite{Kuchar:1990vy,Brown:1994py,Mercuri:2003wn}.
In the present analysis we avoid 
the specification of a particular 
framework, making essentially reference 
to the Schr\"odinger-like dynamics 
of the Universe wave function, because 
on a cosmological setting the physical 
output of an evolutionary approach is 
just the emergence of a new matter 
contribution to the thermal 
bath of the Universe. Such an additional matter 
density defines also a new reference 
frame, which is, to some extent, a 
preferred one, in view of its 
non-test-fluid character.

Hence, it is clear how the question 
concerning the positive value 
of the emerging dust energy density 
(a central theme in the matter-time 
dualism in quantum gravity) is crucial 
when reconstructing the Universe thermal 
history \cite{KolbTurner}.
It is just the request to deal 
with a fully negative spectrum 
of the super-Hamiltonian 
(fully positive dust energy density) 
which leads us to involve a negative 
cosmological constant in the dynamics. 
In so doing we get a contribution in 
the Hamiltonian which corresponds to 
a harmonic oscillator energy, 
but with a global negative sign.
Such a feature has a rather general validity (see Sec. \ref{V}) 
since, even for a generic inhomogeneous 
cosmological model the kinetic 
contribution of the Universe volume 
to the Hamiltonian, together with the negative 
cosmological constant term, provides 
the same harmonic-oscillator structure.  
However, such a contribution implies 
an unstable behavior of the full 
Hamiltonian operator, whose spectrum 
is unbounded from below. 
The regularization is performed via 
a polymer quantum treatment of the 
Universe scale factor (say the Universe 
volume). Such a regularization procedure 
provides a well-defined ground state, 
having the right negative eigenvalue 
for getting a viable phenomenology.
In other words, as proposed in 
\cite{Battisti:2006uj}, we are implementing the 
idea that, on a quantum level, 
it is enough to require that the dust energy 
density corresponding to the ground state be positive
in order to get a reliable physical model.

It is worth stressing how the 
ground state eigenvalue is surprisingly 
unaffected by the value of the negative 
cosmological constant [see \eqref{maxspectpol}] and, therefore, $\Lambda$ is 
freely available for the cosmological 
problem and it can be properly fixed to 
a very small value, whose dynamical 
role will be only to provide a turning point 
in the far future. 
The negative cosmological term, 
here considered, has no ``interference'' 
with the present phenomenon of 
Universe acceleration \cite{Riess:1998cb,Perlmutter:1998np}, 
whose origin could be due to a 
positive cosmological constant 
\cite{Hinshaw:2012aka}, but also to other physical mechanisms 
(as for instance quintessence paradigms \cite{Tsujikawa:2013fta}).
This effect is indeed a late evolutionary 
feature of the Universe, almost 
uncorrelated with the domain of validity 
of this study. Furthermore, 
a positive cosmological constant term 
would not guarantee the spectral 
features traced above.

However, motivations for including 
a negative cosmological constant in 
the Universe dynamics can be found 
in many modern fundamental theories, 
such as supersymmetry \cite{deWit:1999ui} and AdS/CFT correspondence \cite{Beisert:2010jr}.

Finally, we stress how, 
independently of the dust energy density 
origin (a nonvanishing eigenvalue for the super-Hamiltonian 
or a matter clock involved in 
the dynamics) the corresponding cosmological 
evolution could have a deep phenomenological 
impact on the present Universe. We demonstrate that this 
contribution, associated to the 
regularized ground state eigenvalue, 
admits a reliable classical limit 
and then it is redshifted by the 
inflation {\it e}-folding, leaving 
no measurable trace on the observed Universe.

\section{The model}\label{III}

Let us consider a preinflationary Universe described by a flat Friedmann-Robertson-Walker (FRW) metric, whose metric is (we work in units $c=1$)
\begin{equation}
ds^{2}=dt^{2}-a^{2}(t)dl_{RW}^{2},\label{eqn:FRW}
\end{equation}
with
\begin{equation}
dl_{RW}^{2}=dr^{2}+r^{2}(d\theta^{2}+\sin^{2}(\theta)d\phi^{2}),\label{eq:tri_el_lin_FRW}
\end{equation}
and a massive noninteracting scalar field $\phi$, modeling the inflaton trapped into a false vacuum. 
  
The Hamiltonian for such a system in the presence of a negative cosmological constant $-\Lambda$, $\Lambda>0$, reads \footnote{The units are $[\phi]=(\mathrm{length}^{-1})$ and $[m]=\mathrm{mass}$.}
\begin{equation}
\mathcal{H}=-\frac{2\pi G}{3}\frac{p_{a}^{2}}{a}-\frac{\Lambda}{8\pi G} a^{3}+\frac{1}{2\hbar a^{3}}p_{\phi}^{2}+\frac{1}{2}\frac{m^{2}}{\hbar}a^{3}\phi^{2},\label{eq:H_a}
\end{equation}
$p_a$ and $p_\phi$ being the conjugate momenta to $a$ and $\phi$, respectively, while $m$ is the mass of the scalar field. 

In view of quantization a convenient set of phase-space coordinates is obtained via the canonical transformation 
\begin{equation}
a\rightarrow\rho=a^{\nicefrac{3}{2}},\qquad p_a\rightarrow p_{\rho}=\frac{2}{3}p_{a}\rho^{-\nicefrac{1}{3}},\label{eq:def_rho}
\end{equation}
and the Hamiltonian (\ref{eq:H_a}) in the new set of variables takes the following form  
\begin{equation}
\mathcal{H}=-\frac{3\pi G}{2 }p_{\rho}^{2}-\frac{\Lambda}{8\pi G}\rho^{2}+\frac{1}{2\hbar\rho^{2}}p_{\phi}^{2}+\frac{1}2\frac{m^{2}}{\hbar}\rho^{2}\phi^{2}.\label{eq:H_rho}
\end{equation}
The canonical quantization of the associated dynamical system is obtained by replacing the configuration variables $\rho$ and $\phi$ with multiplicative operators defined in a suitable (pre-)Hilbert space and the momenta with the proper derivative operators, {\it i.e.}  
\begin{equation}
p_{\rho}\,\rightarrow\,-i\hbar\frac{\partial}{\partial\rho},\qquad p_{\phi}\,\rightarrow\,-i\hbar \frac{\partial}{\partial\phi},
\end{equation}
such that the Wheeler-DeWitt operator $\hat{\mathcal{H}}$ becomes
\begin{equation}
\hat{\mathcal{H}}\Psi\left(\rho,\,\phi\right)=
\biggl[\frac{3\pi \ell_P^4}{2 G}\frac{\partial^{2}}{\partial\rho^{2}}-\frac{\Lambda}{8\pi G}\rho^{2}-\frac{\hbar}{2\rho^{2}}\frac{\partial^{2}}{\partial\phi^{2}}+\frac{1}2\frac{m^{2}}{\hbar}\rho^{2}\phi^{2}\biggr]\Psi\left(\rho,\,\phi\right),\label{eq:WDW_initial}
\end{equation}
$\ell_P=\sqrt{\hbar G}$ being the Planck length. The evolutionary quantum equation associated
with (\ref{eq:WDW_initial}) reduces to an eigenvalue equation for the
Hamiltonian (\ref{eq:H_rho})
\begin{gather}
\hat{\mathcal{H}}\Psi\left(\rho,\,\phi\right)
=E\Psi\left(\rho,\,\phi\right).\label{eq:WDW-evolutionary}
\end{gather}

\subsection{Quasiclassical limit}
Let us address a Born-Oppenheimer approximation (to be verified {\it a posteriori}), in which we construct the wave function as the product of the following two terms 
\begin{equation}
\Psi\left(\rho,\,\phi\right)=\zeta\left(\rho\right)\chi\left(\rho,\,\phi\right),\label{eq:eigenfunction}
\end{equation}
in which  $\chi$ describes the scalar field wave function and depends parametrically on $\rho$, while $\zeta$ is the Universe wave function. 

Hence, we impose 
\begin{equation}
\left[-\frac{\partial^{2}}{\partial\rho^{2}}+\frac{\Lambda}{12\pi^2 \ell_P^4}\rho^{2}\right]\zeta_{n}\left(\rho\right)=\frac{2G}{3\pi \ell_P^4}E^\rho_n\zeta_{n}\left(\rho\right).\label{eq:osc_arm_zeta}
\end{equation}
The equation above can be written as the Schr\"odinger equation describing a harmonic oscillator with mass $M=\frac{1}{3\pi G}$ and frequency $\Omega=\frac{1}2\sqrt{3\Lambda}$. A solution of the eigenvalue problem (\ref{eq:osc_arm_zeta}) is thus given by
\begin{equation}
\zeta_{n}\left(\rho\right)=N_{n}\Lambda_{\star}^{\nicefrac{1}{8}}e^{-\frac{\sqrt{\Lambda_{\star}}\rho^{2}}{2}}H_{n}\left(\Lambda_{\star}^{\nicefrac{1}{4}}\rho\right),\label{eq:zeta_n}
\end{equation}
where $H_n(x)$ is the Hermite polynomial of degree $n$, $N_n=\sqrt{\frac{1}{\pi^{1/2}\,2^n\, n!}}$ is a normalization constant and
\begin{equation}
\Lambda_{\star}\equiv\frac{\Lambda}{12\pi^2 \ell_P^4}.\label{eq:lambda_star}
\end{equation}
The eigenvalues $E^\rho_n$ reads
\begin{equation}
E_{n}^{\rho}=\hbar \frac{1}{2}\sqrt{3\Lambda}\left(n+\frac{1}{2}\right),\label{eq:autoval_n}
\end{equation}
where $n\in\mathbb{N}$ is restricted to be odd \footnote{
The scale factor $a$ - and consequently $\rho=a^{\nicefrac{3}{2}}$
- cannot take negative values while harmonic oscillator solutions
(\ref{eq:zeta_n}) do admit them. We must then impose that $n$ takes
only odd values in $\mathbb{N}$, corresponding to a wave function
node (\ref{eq:psi}) in $\rho=0$, and consider only the values
$\rho\geq0$.}.

In the same way, we impose 
\begin{equation}
\left[-\frac{\partial^{2}}{\partial\phi^{2}}+\frac{m^{2}}{\hbar^2}\rho^{4}\phi^{2}\right]\chi_{k}\left(\rho,\,\phi\right)=\frac{2\rho^{2}}{\hbar}E^\phi_k\chi_{k}\left(\rho,\,\phi\right),\label{eq:osc_arm_chi}
\end{equation}
whose solution reads
\begin{equation}
\chi_{k}\left(\rho,\,\phi\right)=N_{k}m_\star^{\nicefrac{1}{4}}\sqrt{\rho}e^{-\frac{m_{\star}\phi^{2}\rho^{2}}{2}}H_{k}\left(\sqrt{m_{\star}}\phi\rho\right),\label{eq:chi_k}
\end{equation}
where 
\begin{equation}
m_{\star}\equiv\frac{m}{\hbar},\label{eq:m_star}
\end{equation}
with eigenvalues given by
\begin{equation}
E_{k}^{\phi}=m\left(k+\frac{1}{2}\right),\qquad k\in\mathbb{N}.\label{eq:autoval_k}
\end{equation}
The solution of (\ref{eq:H_rho}) is to be considered in the context
of a theory of small oscillations of the scalar field around the minimum
of its potential. In this way the scalar field cannot explore the
complete profile of the potential $V\left(\phi\right)$, and this
makes the quantum number $k$ confined in a limited interval whose
upper limit depends on the parameters of the model.

In the appendix we demonstrate that an approximated solution of the eigenvalue equation (\ref{eq:WDW-evolutionary}) is given by the product of the functions (\ref{eq:zeta_n}) and
(\ref{eq:chi_k}) for $\rho\rightarrow\infty$, {\it i.e.}
\begin{equation}
\Psi_{n,k}\left(\rho,\phi\right)=\zeta_{n}\left(\rho\right)\chi_{k}\left(\rho,\phi\right)
=N_{n}N_{k}\Lambda_{\star}^{\nicefrac{1}{8}}m_{\star}^{\nicefrac{1}{4}}e^{-\frac{\sqrt{\Lambda_{\star}}\rho^{2}}{2}}H_{n}\left(\Lambda_{\star}^{\nicefrac{1}{4}}\rho\right)\sqrt{\rho}e^{-\frac{m_{\star}\phi^{2}\rho^{2}}{2}}H_{k}\left(\sqrt{m_{\star}}\phi\rho\right),\label{eq:psi}
\end{equation}
with associated eigenvalue 
\begin{align}
E_{n,\, k} & =-E^\rho_n+E^\phi_\rho+\frac{\sqrt{3}}{2}\hbar\sqrt{\Lambda}\left(k+\frac{1}{2}\right)\nonumber \\
 & =\frac{\sqrt{3}}{2}\hbar\sqrt{\Lambda}\left(k-n\right)+m\left(k+\frac{1}2\right).\label{eq:E_{n,k}-2}
\end{align}
Therefore, in the late-time limit $\rho\rightarrow\infty$, the Born-Oppenheimer approximation (\ref{eq:psi}) is well grounded for energy eigenstates. However, the energy spectrum is unbounded from below; thus the resulting dynamical system is unstable.  

Semiclassical states can be constructed through the wave packets peaked around some classical values $\tilde{n}$ and $\tilde{k}$ as follows 
\begin{equation}
\Psi^{[\tilde{n},\tilde{k}]}\left(\rho,\,\phi,\, t\right)=A\,\underset{n=1}{\overset{\infty}{\sum}}\underset{k=1}{\overset{\infty}{\sum}}e^{-\frac{\left(n-\tilde{n}\right)^{2}}{2\sigma_{1}^{2}}}e^{-\frac{\left(k-\tilde{k}\right)^{2}}{2\sigma_{2}^{2}}}\,\zeta_{n}\left(\rho\right)\,\chi_{k}\left(\rho,\phi\right)
e^{-i\frac{E_{n,k} t}{\hbar}},\label{eq:packet_big_rho}
\end{equation} 
where $\sigma_1$ and $\sigma_2$ denote the distribution variances, while $A$ is a normalizing factor.

We want to study the evolution in time of the expectation values of the operator $\rho$ on such states, {\it i.e.} 
\begin{alignat}{1}
\left\langle \rho\right\rangle _{t} & =\left\langle \Psi^{[\tilde{n},\tilde{k}]}(t)\left|\rho\right|\Psi^{[\tilde{n},\tilde{k}]}(t)\right\rangle  =\int_{0}^{\infty}d\rho\int_{-\infty}^{\infty}d\phi\,\left(\Psi^{[\tilde{n},\tilde{k}]}\left(\rho,\,\phi,\, t\right)\right)^*\rho\,\Psi^{[\tilde{n},\tilde{k}]}\left(\rho,\,\phi,\, t\right),
\end{alignat}
and its variation
\begin{align}
\left\langle \Delta\rho^{2}\right\rangle _{t} & =\int_{0}^{\infty}d\rho\int_{-\infty}^{\infty}d\phi\left[\left(\Psi^{[\tilde{n},\tilde{k}]}\left(\rho,\,\phi,\, t\right)\right)^*\rho^{2}\Psi^{[\tilde{n},\tilde{k}]}\left(\rho,\,\phi,\, t\right)\right]-\left\langle \rho\right\rangle _{t}^{2}.
\end{align}
A similar analysis is performed for $\phi$.

\begin{center}
\begin{figure}[H]
\centering
\includegraphics[scale=1]{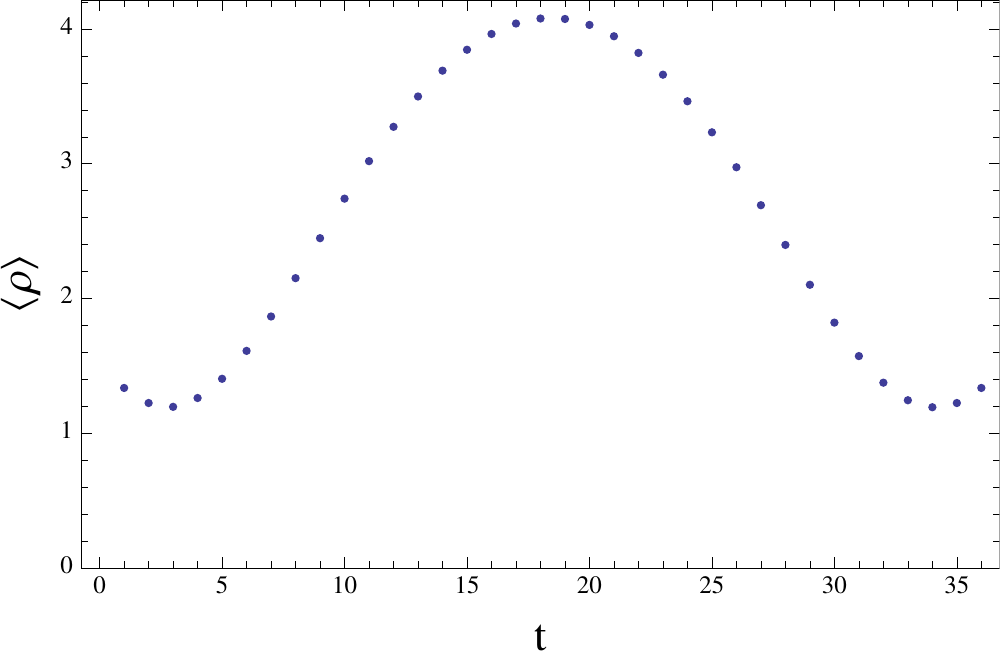}
\caption{
The points denote the expectation value of the scale factor on the wave packet (\ref{eq:packet_big_rho}) 
in units of $\frac{1}{\Lambda_{\star}^{\nicefrac{1}{4}}}$
.\label{fig:rho_t_big rho}}
\end{figure}
\end{center}

\begin{center}
\begin{figure}[H]
\centering
\includegraphics[scale=1]{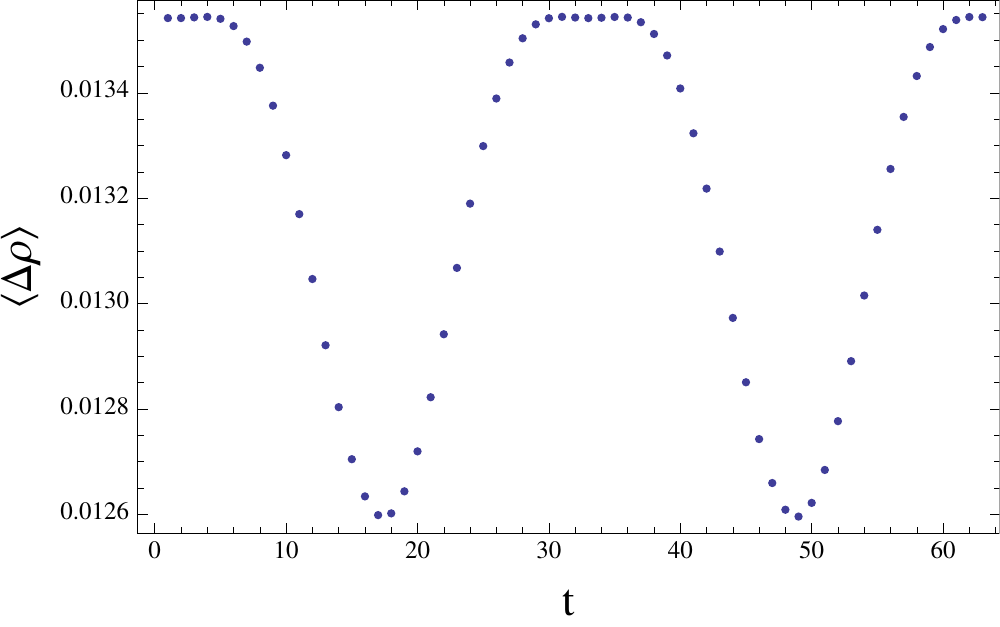}
\caption{The uncertainty $\left\langle \Delta\rho\right\rangle _{t}$
calculated on the wave packet (\ref{eq:packet_big_rho}). $\left\langle \Delta\rho\right\rangle _{t}$
is in units of $\frac{1}{\Lambda_{\star}^{\nicefrac{1}{4}}}$. \label{fig:rho_t_big rho-1-1}}
\end{figure}
\end{center}

\begin{center}
\begin{figure}[H]
\centering
\includegraphics[scale=1]{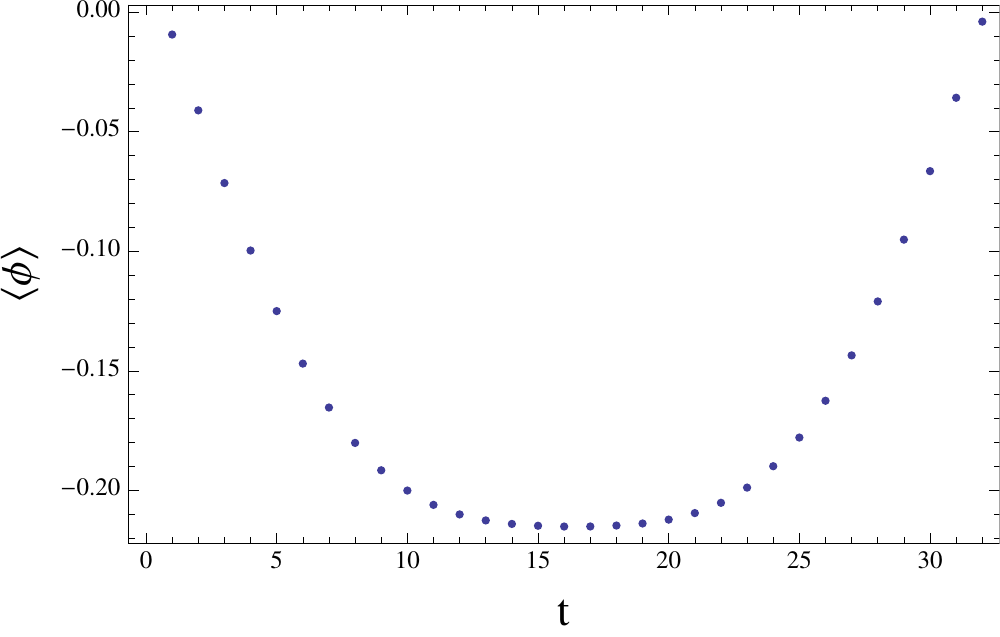}
\caption{The expectation value of the scalar field 
calculated on the wave packet (\ref{eq:packet_big_rho}). $\left\langle \phi\right\rangle _{t}$
is in units of $\frac{\Lambda_{\star}^{\nicefrac{1}{4}}}{\sqrt{m_{\star}}}$.
\label{fig:rho_t_big rho-1-1-1}}
\end{figure}
\end{center}

\begin{center}
\begin{figure}[H]
\centering
\includegraphics[scale=1]{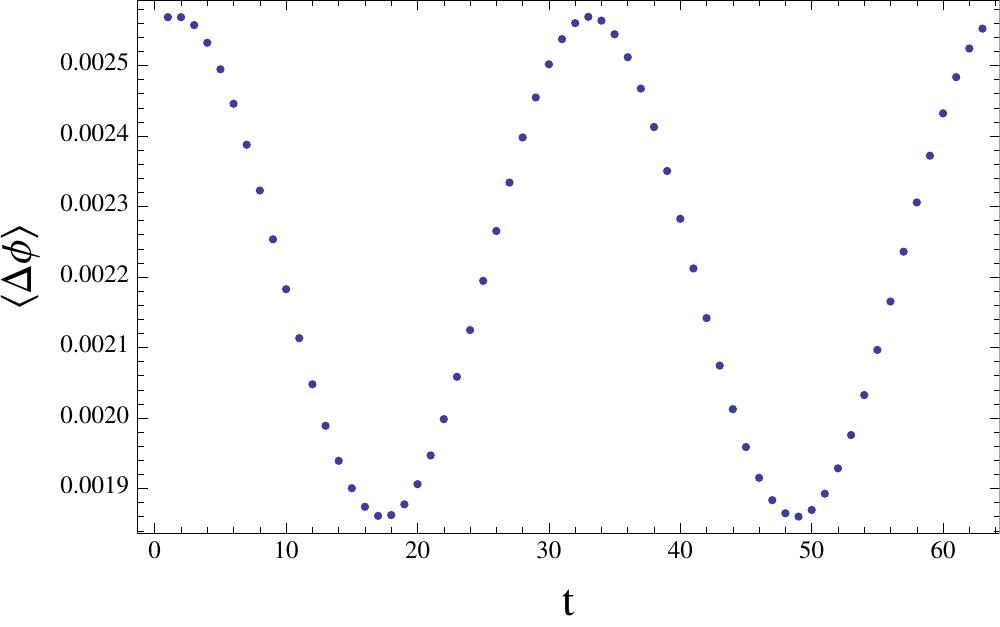}
\caption{Motion of the scalar field uncertainty $\left\langle \Delta\phi\right\rangle _{t}$
calculated on the wave packet (\ref{eq:packet_big_rho}). $\left\langle \Delta\phi\right\rangle _{t}$
is in units of $\frac{\Lambda_{\star}^{\nicefrac{1}{4}}}{\sqrt{m_{\star}}}$.
\label{fig:delta_phi^2 big rho}}
\end{figure}
\end{center}

In Figs. \ref{fig:rho_t_big rho}, \ref{fig:rho_t_big rho-1-1}, \ref{fig:rho_t_big rho-1-1-1} and \ref{fig:delta_phi^2 big rho}, the behaviors of such expectation values and variances are sketched.

We fixed $\tilde{n}=11$, $\tilde{k}=3$, $\sigma_1=0.1$ and $\sigma_2=0.1$, while $\Lambda_{\star}=1$ and $m_{\star}=100$. Time is in units of $\frac{2\pi}{10}\frac{1}{\Omega}=\frac{2\pi}{5\sqrt{3\Lambda}}$.



\noindent 

\noindent 

\noindent 

\noindent 

We see how the variances remain bounded 
and much smaller than the corresponding expectation values. 


\subsection{Close to the singularity}

In the limit $\rho\rightarrow 0$ we expect the potential term of the scalar field in (\ref{eq:H_rho}) to be negligible with respect to the kinetic one, {\it i.e.}
\begin{equation}\label{appr}
\frac{1}{2\rho^{2}}p_{\phi}^{2}\gg\frac{1}2m^{2}\rho^{2}\phi^{2},
\end{equation}
such that the Hamiltonian (\ref{eq:H_rho}) can be rewritten by neglecting this potential term
\begin{equation}\label{eq:H-1}
\mathcal{H}=-\frac{3\pi G}{2 }p_{\rho}^{2}-\frac{\Lambda}{8\pi G}\rho^{2}+\frac{1}{2\hbar\rho^{2}}p_{\phi}^{2}.
\end{equation}
Let us first consider the associated classical system coming out of an evolutionary quantum dynamics, {\it i.e.} for $\mathcal{H}=E$. 
By recalling the expression $p_{a}=-\frac{3c^{2}}{4\pi G}a\dot{a}$, the following Friedmann equation is obtained
\begin{equation}
\left(\frac{\dot{a}}{a}\right)^{2}=\frac{8\pi G}{3}\left(\frac{1}{2\hbar a^{6}}p_{\phi}^{2}-\frac{E}{a^{3}}-\frac{\Lambda}{8\pi G}\right) \label{eq:fried_before}
\end{equation}
where $p_\phi$ is constant since the field is sufficiently frozen near the minimum. An analytic solution is given by 
\begin{equation}
a(t)=\left[A+B\sin\left(\omega t+\phi\right)\right]^{\nicefrac{1}{3}},\label{eq:sol_eq_fried_before}
\end{equation}
where
\begin{gather}
A=-4\pi G\frac{E}{\Lambda},\qquad B=\pm\sqrt{\frac{8\pi G}{\Lambda}\left(\frac{p_{\phi}^{2}}{2\hbar}+2\pi G\frac{E^{2}}{\Lambda}\right)},\qquad\omega=\sqrt{3\Lambda},\label{eq:w,A,B}
\end{gather}
and since $|B|>|A|$ the classical initial singularity $a=0$ is still present. Indeed, it can be shown that there is also a final \textit{big crunch} singularity, as usual in models with a negative cosmological constant.

The scalar field dynamics can also be solved analytically, finding 
\begin{equation}
\phi=\frac{p_{\phi}}{\hbar\omega}\frac{2}{\sqrt{B^{2}-A^{2}}}\textrm{arctanh}\Bigg(\frac{B}{\sqrt{B^{2}-A^{2}}}
+\frac{A}{\sqrt{B^{2}-A^{2}}}\,\tan\left(\frac{1}{2}\left(\omega t+\phi\right)\right)\Bigg)+\textrm{cost.}
\end{equation}

\begin{center}
\begin{figure}[H]
\centering
\includegraphics{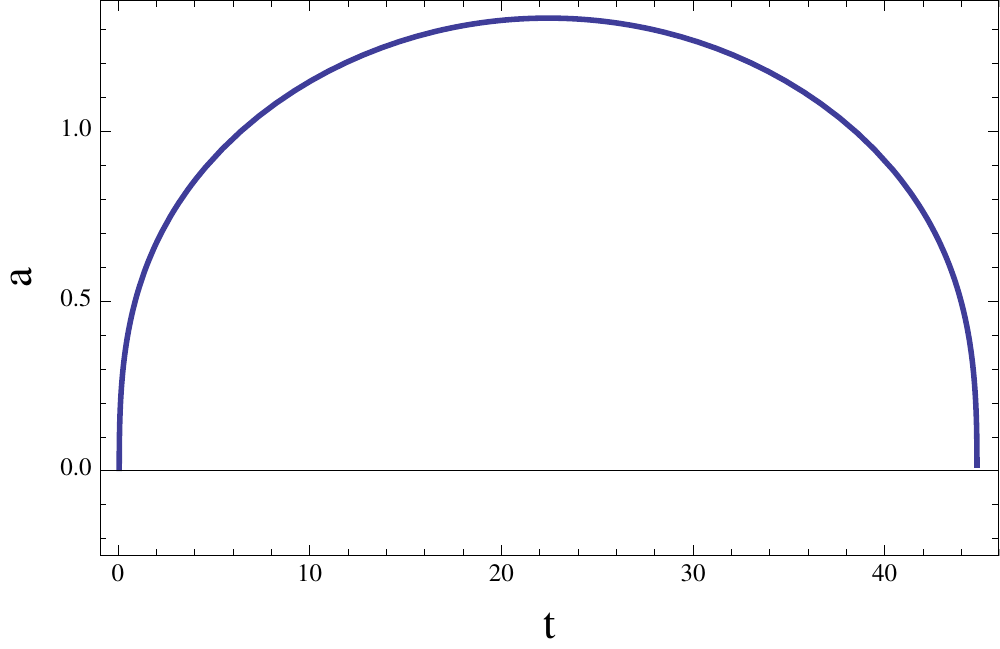}
\caption{Classical trajectory of the scale factor $a$ in function
of time. Time is in units of $\frac{2\pi}{10\omega}$ and the scale factor
$a\left(t\right)$ in units of $B^{\nicefrac{1}{3}}$. It is worth noting how the singularity $a=0$ is reached in a finite amount of time both in the past and in the future.} 
\label{aclass}
\end{figure}
\end{center}

\begin{center}
\begin{figure}[H]
\centering
\includegraphics{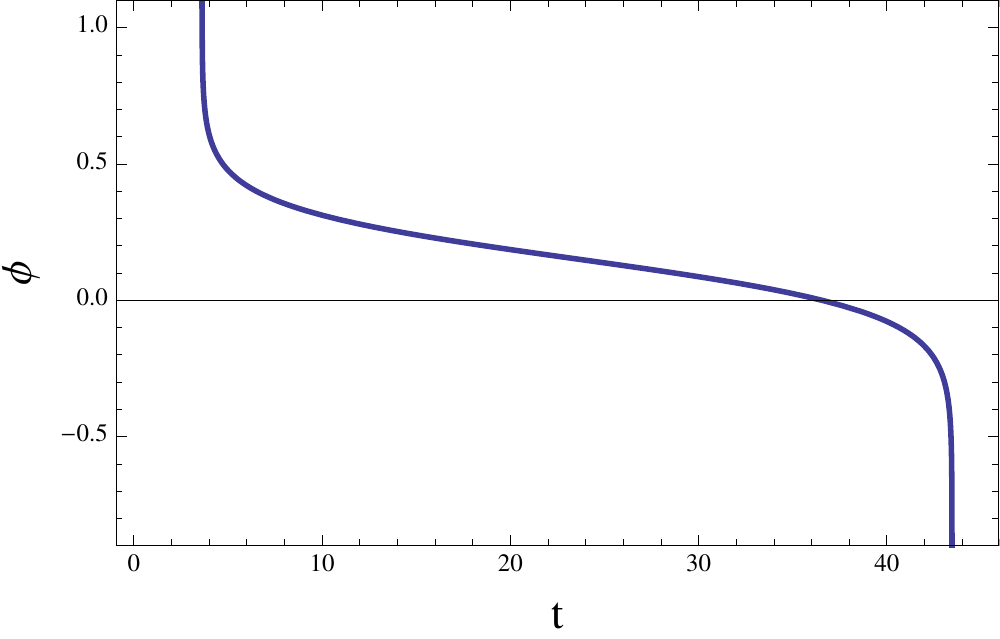}
\caption{Classical trajectory of $\phi$ in function of time.
Time is in units of $\frac{2\pi}{10\omega}$, the scalar field in units
of $\frac{p_{\phi}}{\hbar\omega}\frac{2}{\sqrt{B^{2}-A^{2}}}$.}
\label{phiclass}
\end{figure}
\end{center}

The classical behavior of $a$ and $\phi$ is depicted in Figs. \ref{aclass} and \ref{phiclass}.

\noindent 

\noindent 

It is worth noting that the presence of a negative cosmological constant ensures the existence of a turning point in the Universe late classical evolution. 

The associated quantum equation in evolutionary quantum gravity reads
\begin{equation}
\biggl[\frac{3\pi \ell_P^4}{2 G}\frac{\partial^{2}}{\partial\rho^{2}}-\frac{\Lambda}{8\pi G}\rho^{2}-\frac{\hbar}{2\rho^{2}}\frac{\partial^{2}}{\partial\phi^{2}}\biggr]\Psi\left(\rho,\,\phi\right)=E\,\Psi\left(\rho,\,\phi\right),\label{eq:WDW-1}
\end{equation}
We look for a plane-wave solution for the scalar field, {\it i.e.} 
\begin{equation}
\Psi\left(\rho,\,\phi\right)=\frac{1}{\sqrt{2\pi}}e^{i\frac{k_{\phi}\phi}{\hbar}}\,\zeta\left(\rho\right),
\end{equation}
such that one gets the following equation for $\zeta$  
\begin{equation}
\biggl[\frac{\partial^{2}}{\partial\rho^{2}}-\Lambda_{\star}\rho^{2}+\frac{k_{\phi\star}^{2}}{\rho^{2}}\biggr]\zeta\left(\rho\right)=E_{\star}\,\zeta\left(\rho\right).\label{eq:small_phi_eq_zeta}
\end{equation}
with  
\begin{equation}
k_{\phi\star}^{2}=\frac{G}{3\pi\hbar \ell_P^4}\,k_\phi^2\, ,\qquad E_{\star}=\frac{2G}{3\pi \ell_P^4}\, E.\label{eq:def_star}
\end{equation}
A solution of Eq.(\ref{eq:small_phi_eq_zeta}) can be constructed by considering the following wave function 
\begin{equation}
\zeta=f\left(\rho\right)e^{\frac{-\sqrt{\Lambda_{\star}}\rho^{2}}{2}}\rho^{\lambda},\label{eq:trial_sol_prod}
\end{equation}
where \footnote{The two functions $e^{\frac{-\sqrt{\Lambda_{\star}}\rho^{2}}{2}}$ and $\rho^{\lambda}$ entering (\ref{eq:trial_sol_prod}) are the solutions of the associated Wheeler-DeWitt equation in the two limits $\rho\rightarrow \infty$ and $\rho\rightarrow 0$.} 
\begin{equation}
\lambda=\frac{1+\sqrt{1-4\, k_{\phi\star}^{2}}}{2}.
\end{equation}
In what follows, we consider the solution with real $\lambda$'s, {\it i.e.}  
\begin{equation}
|k_{\phi\star}|<\frac{1}{2}\label{eq:k_phi<<1}.
\end{equation}
We assume a power series expansion for the unknown function $f(\rho)$ 
\begin{equation}
f\left(\rho\right)=\sum_{n=0}^{n'}c_{n,n'}\rho^{n},\qquad n,n'\in 2\mathbb{N},\label{eq:f(rho)_powers}
\end{equation}
and the eigenvalue problem (\ref{eq:trial_sol_prod}) provides the following difference equation 
\begin{equation}
c_{n+2,n'}(n+2)\left(\left(\sqrt{1-4k_{\phi\star}^{2}}+n+2\right)\right)
-c_{n,n'}\left(E_{\star}+\sqrt{\Lambda_{\star}}\left(\sqrt{1-4k_{\phi\star}^{2}}+2n+2\right)\right)=0\,, \label{eq:eq_coeff_c_n}
\end{equation}
with the restriction 
\begin{equation}
E_{\star}=-\sqrt{\Lambda_{\star}}\left(\sqrt{1-4k_{\phi\star}^{2}}+2n'+2\right).
\end{equation}
The expression of the coefficients $c_{n,n'}$ can be written in terms of (the analytically continued) Euler $\Gamma$-function as follows \cite{Nouredine} 
\begin{equation}
c^{sol}_{n,n'}=\frac{\left(\left(-1\right)^{n}+1\right)\Gamma\left(1+\frac{1}{2}\sqrt{1-4k_{\phi\star}^{2}}\right)\Lambda_{\star}^{n/4}}{\Gamma\left(\frac{n}{2}+1\right)\Gamma\left(\frac{n}{2}+\frac{1}{2}\sqrt{1-4k_{\phi\star}^{2}}+1\right)}\left(-1\right)^{\frac{n}{2}}\frac{\frac{n'}{2}!}{\left(\frac{n'}{2}-\frac{n}{2}\right)!}.\label{eq:c_n}
\end{equation}
and the solution of Eq. (\ref{eq:WDW-1})
thus reads 
\begin{equation}
\Psi_{n',\, k_{\phi}}(\rho,\,\phi)=C\,e^{i\frac{k_{\phi}\,\phi}{\hbar}}\, e^{\frac{-\sqrt{\Lambda_{\star}}\rho^{2}}{2}}\rho^{\frac{1+\sqrt{1-4k_{\phi\star}^{2}}}{2}}\sum_{n=0}^{n'}c_{n,n'}^{sol}\rho^{n}.\label{eq:autofunz_k_phi<<1}
\end{equation}
where $C$ is a normalizing factor and with energy eigenvalue
\begin{equation}
E_{n',\, k_{\phi}}=-\hbar\frac{\sqrt{3\Lambda}}{4}\left(\sqrt{1-4k_{\phi\star}^{2}}+2n'+2\right).\label{eq:E_condiz_3-1-1}
\end{equation}
Indeed, the procedure we used to solve Eq. (\ref{eq:small_phi_eq_zeta}) is just the generalization of those adopted for a three-dimensional harmonic oscillator, in which case $k_{\phi\star}^{2}=-l(l+1)$, 
$l$ being the angular momentum. In fact, the eigenfunction (\ref{eq:autofunz_k_phi<<1}) and the eigenvalue (\ref{eq:E_condiz_3-1-1})
correspond to the three-dimensional generalization of the harmonic
oscillator eigenfunctions and eigenvalues. In particular, the sum $\sum_{n=0}^{n'}c_{n}^{sol}\rho^{n}$
coincides with Laguerre polynomials 
generalized for continuous $k_{\phi}$ values. Within this model, the value of $k_{\phi\star}$ determines if there is a physical singularity \footnote{In quantum mechanics the singularity takes the significance of finding the particle exactly localized in the source, phenomenon
also known as ``fall to the center''.} in $\rho=0$: it is absent for $k_{\phi\star}<\frac{1}{2}$ and present for $k_{\phi\star}>\frac{1}{2}$ 
\cite{Calogero:1969xj,LL Quantum Mechanics}.
Therefore, by assuming the condition (\ref{eq:k_phi<<1}), we get a nonsingular scenario for the early Universe and the big-bang singularity is removed.

\begin{center}
\begin{figure}[H]
\centering
\includegraphics[scale=1]{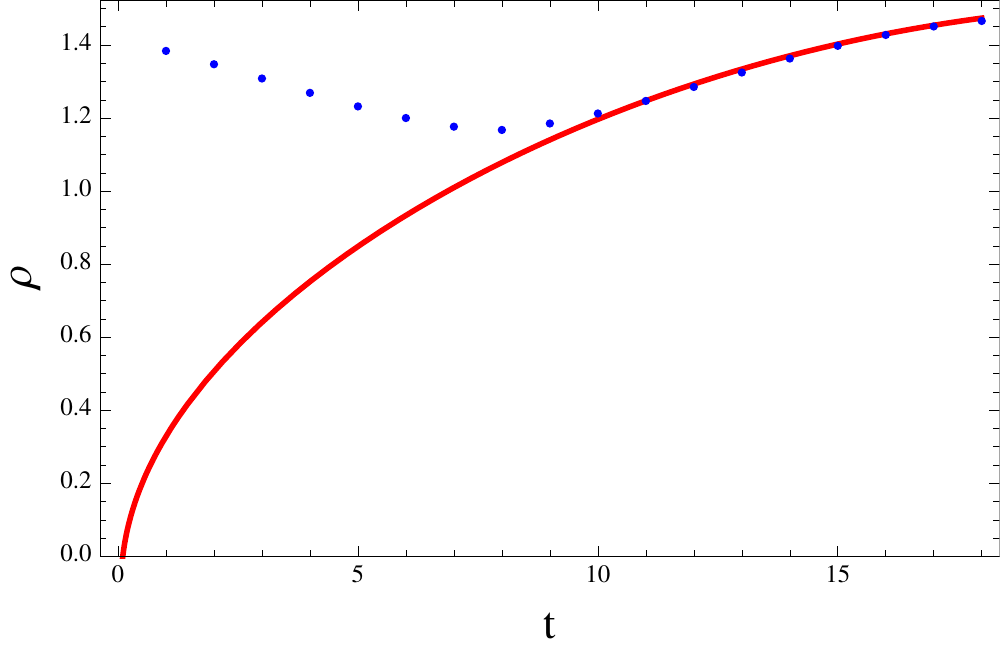}
\caption{The points denote the expectation value of the scale factor on the wave packet (\ref{eq:packet_small_phi}), while the continuous line is the classical behavior of $\rho$ (both in units of $\frac{1}{\Lambda_{\star}^{\nicefrac{1}{4}}}$). We note how the initial singularity is avoided in the quantum model. 
\label{fig:rho_t_big rho-2}}
\end{figure}
\end{center}

Semiclassical states can now be constructed as follows 
\begin{equation}
\Psi^{[\tilde{k}_\phi,\tilde{n}]}\left(\rho,\,\phi,\, t\right)=A\, \intop_{-\nicefrac{1}{2}}^{+\nicefrac{1}{2}}dk_{\phi\star}\,e^{-\frac{\left(k_{\phi}-\tilde{k}_{\phi}\right)^{2}}{2\alpha^{2}}}\underset{n=1}{\overset{\infty}{\sum}}e^{-\frac{\left(n-\tilde{n}\right)^{2}}{2\sigma^{2}}}
\,e^{-i\frac{E_{n,\, k_{\phi}}t}{\hbar}} \Psi_{n,\, k_{\phi}}(\rho,\phi).\label{eq:packet_small_phi}
\end{equation} 
$A$ being a normalizing factor. The behavior of expectation values and distribution variances is presented in Figs. \ref{fig:rho_t_big rho-2} and \ref{fig:rho_t_big rho-2-1-1} for $\sigma=0.1$.

\noindent 

\noindent \begin{center}
{\footnotesize }
\begin{figure}[H]
\centering
\includegraphics[scale=1]{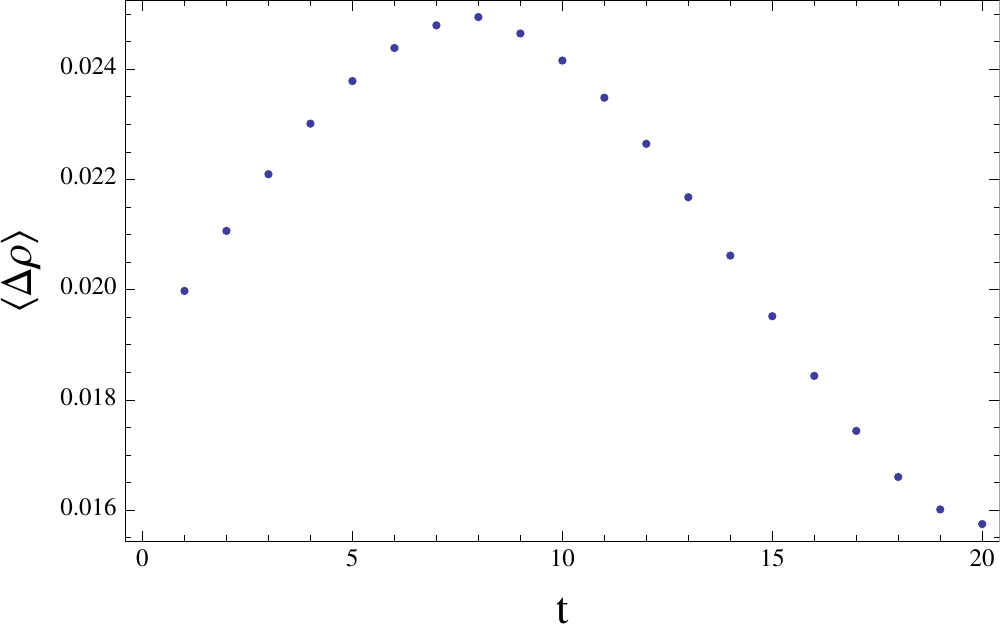}
\caption{The uncertainty of $\rho$ on the wave packet (\ref{eq:packet_small_phi}). $\left\langle \Delta\rho\right\rangle _{t}$
is in units of $\frac{1}{\Lambda_{\star}^{\nicefrac{1}{4}}}$. \label{fig:rho_t_big rho-2-1-1}}
\end{figure}
\end{center}

This analysis confirms that the quantum model is nonsingular as $\rho\rightarrow 0$, since  the initial singularity is replaced by a bounce. Furthermore, deviation becomes smaller and smaller as $t$ increases; thus the Universe becomes more and more classical at late times.

\section{Polymer quantization}\label{IV}

In this section we implement polymer quantization for the variable $\rho$ describing the Universe volume. We will outline how this procedure provides a bounded-from-below dust energy contribution, which allows us to fix the value of $E$ owing to the relaxation of the Universe to the fundamental state. 

Polymer quantization \cite{Ashtekar:2002sn,Corichi:2007tf} realizes a representation of the Weyl algebra which is not unitary equivalent to the Schr\"odinger representation. In fact, it is based on violating one of the hypotheses of the Stone-von Neumann uniqueness theorem, namely weak-continuity. This is due to the fact that the space of configuration variables is endowed with a discrete topology, such that only finite translations can be implemented and momenta are not defined. Hence, in order to write a proper Hamiltonian, one is forced to introduce a lattice in the configuration space and to approximate the momenta by the action of the translation operator on the minimum lattice distance (polymer scale). In this sense, polymer quantization naturally accounts for the presence of a fundamental discrete structure, as expected in quantum gravity approaches. As a consequence, the quantum dynamics is significantly affected at  scales which are comparable with the polymer scale, while in the continuum limit there is a substantial overlap with the results obtained in the Schr\"odinger representation. 

Let us address polymer quantization for $\rho$. The associated wave function in the frozen case
can be written as 
\begin{equation}
\Psi^{\scriptscriptstyle{pol}}\left(\rho,\,\phi\right)=\zeta^{\scriptscriptstyle{pol}}\left(\rho\right)e^{i\frac{k_{\phi}\phi}{\hbar}},
\end{equation}
where the superscript $^{\scriptscriptstyle{pol}}$ denotes the polymer part of the wave function. The function $\zeta^{\scriptscriptstyle{pol}}$ belongs to the Hilbert space of square-integrable functions over the Bohr compactification of the real line. It can be expanded on the eigenvectors of the polymer operator $\rho$, whose spectrum is defined on the lattice $\mathcal{L}_{\lambda_\star}=\{\rho=n\lambda_\star,\, n\in \mathbb{Z}\}$, {\it i.e.}
\begin{equation}    
\hat{\rho}\,|n\lambda_\star\rangle=n\lambda_\star|n\lambda_\star\rangle,
\end{equation}
$\lambda_\star$ being the polymer scale. Finite translations act as 
\begin{equation}
T_\rho\, |\rho'\rangle=|\rho'+\rho\rangle,
\end{equation}
and the momentum operator can be defined in terms of them as follows 
\begin{equation}
\hat{p}_\rho^{\scriptscriptstyle{pol}}= \frac{\hbar}{2i\lambda_\star}(T_{\lambda_\star}- T_{-\lambda_\star}),
\end{equation} 
which formally coincides with the replacement 
\begin{equation}
\hat{p}_\rho\rightarrow \frac{1}{\lambda_\star}\sin{\left(\frac{\lambda_\star \hat{p}_\rho}{\hbar}\right)}. 
\end{equation}
The eigenvalue problem (\ref{eq:small_phi_eq_zeta}) becomes
\begin{equation}
\left(\widehat{H}^{\scriptscriptstyle{PHO}}+k_{\phi}^{2}\widehat{\frac{1}{\rho^{2}}}\right)\zeta^{\scriptscriptstyle{pol}}\left(\rho\right)=-E\zeta^{\scriptscriptstyle{pol}}\left(\rho\right),\label{eq:PQM-H}
\end{equation}
where $\widehat{H}^{\scriptscriptstyle{PHO}}$ is the Hamiltonian operator for the harmonic oscillator in the polymer representation, which in the momentum polarization reads \cite{Hossain:2010eb} 
\begin{equation}
\widehat{H}^{\scriptscriptstyle{PHO}}\tilde{\zeta}^{\scriptscriptstyle{pol}}(p_\rho)=\frac{\hbar^2}{8M\lambda^2_{\star}}\left[2-2\cos\left(\frac{2\lambda_{\star}p_\rho}{\hbar}\right)\right]\tilde{\zeta}^{\scriptscriptstyle{pol}}(p_\rho)-\frac{\hbar^2}{2}M\Omega^{2}\frac{d^{2}\tilde{\zeta}^{\scriptscriptstyle{pol}}(p_\rho)}{dp_\rho^{2}},
\end{equation}
where $M$ and $\Omega$ denote the mass and the frequency of the harmonic oscillator and they take the following expression in terms of the parameters of the model
\begin{equation}
M=\frac{1}{3\pi G},\qquad\Omega=\frac{\sqrt{3\Lambda}}{2}.
\end{equation}
Let us assume that the second term on the left-hand side of (\ref{eq:PQM-H}) is a small perturbation and let us solve first the eigenvalue problem for the polymer harmonic oscillator: 
\begin{equation}
\frac{\hbar^2}{8M\lambda_{\star}}\left[2-2\cos\left(\frac{2\lambda_{\star}p_\rho}{\hbar}\right)\right]\tilde{\zeta}^{\scriptscriptstyle{pol}}-\frac{\hbar^2}{2}M\Omega^{2}\frac{d^{2}\tilde{\zeta}^{\scriptscriptstyle{pol}}}{dp_\rho^{2}}=-E\tilde{\zeta}^{\scriptscriptstyle{pol}}.\label{eq:eq_SHO_polymer}
\end{equation}
By introducing the following quantities 
\begin{equation}
\quad u=\frac{\lambda_{\star}p_\rho}{\hbar}+\frac{\pi}{2},\quad\alpha=-\frac{2E}{\hbar\Omega g}-\frac{1}{2g^{2}},\quad g=\frac{M\Omega\lambda_{\star}^{2}}{\hbar},
\end{equation}
Eq. (\ref{eq:eq_SHO_polymer}) takes the form of Mathieu
equation
\begin{equation}
\frac{d^{2}\tilde{\zeta}^{\scriptscriptstyle{pol}}}{du^{2}}+\left[\alpha-\frac{1}{2}g^{-2}\cos\left(2u\right)\right]\tilde{\zeta}^{\scriptscriptstyle{pol}}=0.
\end{equation}
We restrict to those solutions which are periodic (or antiperiodic) in $u$, since the associated conjugate variable $\rho$ is discrete. These solutions of the Mathieu equation are parametrized by $g$ and they can be written as  
\begin{align}
\tilde{\zeta}^{\scriptscriptstyle{pol}}_{2n}\left(u\right) & =\pi^{-\nicefrac{1}{2}}\textrm{ce}_{n}\left(u,g\right),\qquad\alpha=A_{n}\left(g\right)\label{matce} \\
\tilde{\zeta}^{\scriptscriptstyle{pol}}_{2n+1}\left(u\right) & =\pi^{-\nicefrac{1}{2}}\textrm{se}_{n}\left(u,g\right),\qquad\alpha=B_{n}\left(g\right)\label{matse}
\end{align}
where $\textrm{ce}_{n}$ and $\textrm{se}_{n}$ $\left(n\in\mathbb{N}\right)$
are respectively the sine and cosine elliptic functions, while $A_{n}$
and $B_{n}$ are the \textit{characteristic values
}\textit{functions}. For even $n$,
$\textrm{ce}_{n}$ and $\textrm{se}_{n}$ are $\pi-$periodic, while
for odd $n$ they are $\pi-$antiperiodic. The energy eigenvalues are
\begin{align}
E_{2n} & =-\frac{\hbar\Omega g}{2}\left(A_{n}\left(g\right)+\frac{1}{2g^2}\right),\label{eq:PSHOeigen1}\\
E_{2n+1} & =-\frac{\hbar\Omega g}{2}\left(B_{n}\left(g\right)+\frac{1}{2g^2}\right).\label{eq:PSHOeigen2}
\end{align}
It is worth noting how in our case the condition $g\ll 1$ holds, since we have
\begin{equation}
g=\lambda_\star^2\frac{\sqrt{3\Lambda}}{3\pi \ell_P^2}\ll 1\rightarrow \Lambda\ll \frac{3\pi^2}{\ell_P^2},
\end{equation}
where we fixed the polymer scale $\lambda_\star\sim \ell_P^{3/2}$. The condition above is an intrinsic consistency condition for our model, since if it was violated then the negative cosmological constant would drive the Universe evolution from the Planck time up to now. Hence, we can expand the characteristic values in powers of $g$ near $g=0$, so getting 
\begin{equation}
\alpha=(2n+\frac{1}{2})g^{-1}-\frac{2n^2+2n+1}{4}+O(g),
\end{equation}
such that the energy spectrum becomes
\begin{equation}\label{energypolymer}
E_n=-\frac{\hbar\sqrt{3\Lambda}}{2}\left(\left(n+\frac{1}{2}\right)-\frac{2n^2+2n+1}{8}g+O\left(g^2\right)\right).
\end{equation}
It is worth noting how the spectrum exhibits a minimum corresponding to 
\begin{equation}\label{maxspectpol}
n_{\scriptscriptstyle{min}}\sim \frac{2}{g}\rightarrow E_{\scriptscriptstyle{min}}\sim
\frac{\hbar\sqrt{3\Lambda}}{2g}=\frac{3\pi\hbar\ell_P^2}{2\lambda_\star^2}.
\end{equation}
Therefore, the dynamical system is now endowed with a ground state, corresponding to the minimum energy eigenvalue, and one can assume the Universe to dynamically relax into such a state. Furthermore, the minimum eigenvalue is negative and the dual dust field energy density is positive; thus it behaves as an ordinary dust field contribution.

Let us now estimate the additional term in (\ref{eq:PQM-H}). By performing the Fourier transform of Mathieu functions it is possible to describe the eigenfunctions (\ref{matce}) and (\ref{matse}) in the coordinate representation. We consider only the Mathieu sine periodic functions, since the same estimate can be repeated for the cosine, and with odd index, since nothing changes for an even one. Hence, $\textrm{se}_{2n+1}$ can be expanded as follows 
\begin{align}
\textrm{se}_{2n+1}\left(u,g\right) & =\sum_{m=0}^{\infty}B_{2m+1}^{2n+1}\left(g\right)\:\sin[(2m+1)u],\label{eq:se_n dispari},
\end{align}
where $B_{2m+1}^{2n+1}\left(g\right)$ are Fourier coefficients for which 
\begin{equation}\label{compl}
\sum_{m=0}^{\infty}|B_{2m+1}^{2n+1}|^2=1.
\end{equation}
The normalized energy eigenstates $|E_{2n+1}\rangle$, for which $\langle p_\rho|E_{2n+1}\rangle=\tilde{\zeta}^{\scriptscriptstyle{pol}}_{2n+1}$, can thus be expanded in terms of $|\rho\rangle=e^{i\rho p_\rho/\hbar} |p_\rho\rangle$ as follows:
\begin{equation}   
|E_{2n+1}\rangle=\frac{1}{\sqrt{2}}\sum_{m=-\infty}^{+\infty} (-)^m\,B^{2n+1}_{|2m+1|}\,|(2m+1)\lambda_\star\rangle.
\end{equation} 
Inverse powers of $\rho$ can be regularized in polymer representation using the expression \cite{Husain:2007bj} 
\begin{equation}
\widehat{\frac{\textrm{sgn}\left(\rho\right)}{\sqrt{\left|\rho\right|}}}\rightarrow\frac{1}{\lambda_{\star}}\left(\sqrt{\left|\rho+\lambda_\star\right|}-\sqrt{\left|\rho-\lambda_\star\right|}\right),\label{eq:sgn(x)/sqrt(x)}
\end{equation}
which provides for $1/\rho^2$
\begin{equation}
\widehat{\frac{1}{\rho^{2}}}\rightarrow\frac{1}{\lambda^4_{\star}}\left(\sqrt{\left|\rho+\lambda_\star\right|}-\sqrt{\left|\rho-\lambda_\star\right|}\right)^4.
\end{equation}
Hence, the expectation value of the additional term in (\ref{eq:PQM-H}) on energy eigenstates reads
\begin{equation}\label{deltah}
k_{\phi}^{2}\,\left\langle E_{2n+1}\left|\widehat{\frac{1}{\rho^{2}}}\right|E_{2n+1}\right\rangle=\frac{k_{\phi}^{2}}{\lambda^2_{\star}}\sum_{m=0}^{\infty}|B_{2m+1}^{2n+1}|^2\left(\sqrt{\left|2m+2\right|}-\sqrt{\left|2m\right|}\right)^4\leq \frac{k_{\phi}^{2}}{\lambda^2_{\star}}\sum_{m=0}^{\infty}\left(\sqrt{\left|2m+2\right|}-\sqrt{\left|2m\right|}\right)^4,
\end{equation}
where we used the condition (\ref{compl}). Let us note that 
\begin{equation}
\left(\sqrt{n+1}-\sqrt{n-1}\right)^{4}<\frac{2}{n^{2}}\quad\forall\: n>1,
\end{equation}
and thus
\begin{equation}
\sum_{m=0}^{\infty}\left(\sqrt{\left|2m+2\right|}-\sqrt{\left|2m\right|}\right)^4<
4+\sum_{m=1}^{\infty}\frac{2}{(2m+1)^{2}} = 2+\sum_{m=0}^{\infty}\frac{2}{(2m+1)^{2}}=
2+\frac{\pi^{2}}{4},\label{summ}
\end{equation}
such that for the expression (\ref{deltah}) one has  
\begin{equation}
k_{\phi}^{2}\,\left\langle E_{2n+1}\left|\widehat{\frac{1}{\rho^{2}}}\right|E_{2n+1}\right\rangle<
\frac{8+\pi^{2}}{4}\frac{k_{\phi}^{2}}{\lambda^2_{\star}}.
\end{equation}
The ratio $\Delta$ of this term with the ground state energy eigenvalue reads
\begin{equation}
\Delta=\frac{k_{\phi}^{2}\,\left\langle E_{2n+1}\left|\widehat{\frac{1}{\rho^{2}}}\right|E_{2n+1}\right\rangle}{E_{\scriptscriptstyle{min}}}< \frac{8+\pi^{2}}{4}\,k^2_{\phi\star},
\end{equation}
and it tells us if the additional term is actually a small perturbation or not. For instance, we can require $\Delta<10^{-1}$, so getting   
\begin{equation}
k_{\phi\star}<0.15.
\end{equation}

We see how the smaller the value of $k_{\phi\star}$, the more accurate the energy spectrum \eqref{energypolymer} we got in polymer representation.

\section{Conclusion}\label{V}

We presented a self-consistent cosmological
picture, based on the implementation of
an evolutionary quantum gravity approach to
a reliable model of the primordial isotropic
Universe. Indeed, we construct a nonsingular
cosmology, corresponding to a cyclic Universe,
having a quantum big bounce in the past,
associated to the details of its quantum dynamics,
and a late turning point, due to the presence of
a small negative cosmological constant.
The main issue of the analysis above consists
in determining, via a polymer quantum
approach, a ground state of the Universe,
associated to a positive
dust energy density and to a high occupation number.
For such a vacuum state of the model, it is
possible to construct a reliable classical limit,
which reconciles the early evolutionary Universe
phase with a standard preinflationary scenario (with a decay from a false vacuum).
Furthermore, the emerging dust energy density is
strongly redshifted by the Universe de Sitter phase
and thus, the postinflationary dynamics is 
indistinguishable from the Standard Model one.
Finally, we observe that the value of the negative
cosmological constant does not enter the
ground state energy of the Universe and so it
can be easily stated as an amount which is
unable to affect the present Universe dynamics.
We can consequently conclude that an evolutionary
quantum cosmology exists, able to solve the
singular nature of the big bang, without
any dynamical discrepancy with respect to
a standard Friedmann Universe and with the
additional feature of a natural picture of the
model classical limit.

Despite the fact that the present study offers an interesting cosmological paradigm for the matter-time dualism and the related evolutionary quantum gravity, we stress how the problem of a unique definition of a physical clock for the dynamics of the gravitational field in canonical quantum gravity remains one of the most challenging topics in this area and deserves attention for both its fundamental developments and specific applications. 

We conclude by observing how the
relevance of the present model
must be also recognized in its
possibly very general character.
In fact, the anisotropic Universe degrees
of freedom are dynamically equivalent
to a scalar field and the associated
potential (due to the cosmological model
spatial curvature) is quadratic in the
limit of a quasi-isotropic Universe.
Then, we could first extend the
present model to the homogeneous
Bianchi Universe
(in particular to the Bianchi IX
model, generalizing the closed isotropic
Universe) and then to a generic inhomogeneous
Universe, via a quantum version of
the so-called Belinski-Lipschitz-Kalatnikov conjecture \cite{Belinsky:1982pk,Kirillov:1997fx}.

Although, differently from the scalar
field dynamics, there is no firm
evidence that the limit of small
anisotropies is prescribed by the
Universe evolution, the perspective
to replace the scalar field with
the cosmological gravitational field
degrees of freedom, opens very general
and intriguing scenarios for the
implementation of the idea traced here.

\section{Appendix: Born-Oppenheimer approximation}

We demonstrate that the solutions of the evolutionary equation \eqref{eq:WDW-evolutionary} in the large $\rho$ limit can be constructed as the product of the two harmonic oscillator solutions \eqref{eq:zeta_n} and \eqref{eq:chi_k}. In particular, by inserting the expression \eqref{eq:psi} into Eq. \eqref{eq:WDW-evolutionary} one gets 
\begin{align}
&\biggl[\frac{3\pi \ell_P^4}{2 G}\frac{\partial^{2}}{\partial\rho^{2}}-\frac{\Lambda}{8\pi G}\rho^{2}-\frac{\hbar}{2\rho^{2}}\frac{\partial^{2}}{\partial\phi^{2}}+\frac{1}2\frac{m^{2}}{\hbar}\rho^{2}\phi^{2}\biggr]\zeta_{n}\left(\rho\right)\chi_{k}\left(\rho,\,\phi\right)\nonumber\\
&=\frac{3\pi \ell_P^4}{2 G}\left\{ \zeta''_{n}\left(\rho\right)\chi_{k}\left(\rho,\,\phi\right)+\zeta{}_{n}\left(\rho\right)\chi''_{k}\left(\rho,\,\phi\right)+2\zeta'{}_{n}\left(\rho\right)\chi'_{k}\left(\rho,\,\phi\right)\right\} \nonumber\\
&-\frac{\hbar}{2\rho^{2}}\zeta_{n}\left(\rho\right)\frac{\partial^{2}}{\partial\phi^{2}}\chi_{k}\left(\rho,\,\phi\right)+\left(\frac{1}2\frac{m^2}{\hbar}\rho^{2}\phi^{2}-\frac{3\pi \ell_P^4}{2 G}\Lambda_{\star}\rho^{2}\right)\zeta_{n}\left(\rho\right)\chi_{k}\left(\rho,\,\phi\right)\nonumber\\
&=E\zeta_{n}\left(\rho\right)\chi_{k}\left(\rho,\,\phi\right),\label{eq:WDW_chi'',zeta'',2chi'zeta'-1}
\end{align}
where $'$ denotes the derivative with respect to the
variable $\rho$. These derivatives can be evaluated thanks to the following identity for Hermite polynomials
\begin{equation}
\frac{\partial}{\partial\rho}H{}_{n}\left(\rho\right)=2nH_{n-1}\left(\rho\right),\label{eq:rel_der_pol_Herm}
\end{equation}
and the following asymptotic recurrence relation holding as soon as $\rho\rightarrow\infty$ 
\begin{equation}
H_{n-1}\left(\rho\right)\simeq\frac{1}{2\rho}H_{n}\left(\rho\right),
\end{equation}
which can be combined together, so getting 
\begin{equation}
\frac{\partial}{\partial\rho}H{}_{n}\left(\rho\right)\simeq\frac{n}{\rho}H_{n}\left(\rho\right).\label{eq:rel_der_Herm_inf}
\end{equation}
Hence, $\zeta'_{n}(\rho)$ in the limit $\rho\rightarrow\infty$ reads
\begin{align}
\zeta'_{n}(\rho) & \simeq\left[-\sqrt{\Lambda_{\star}}\rho+\frac{n}{\rho}\right]\zeta_{n}\left(\rho\right),\label{eq:zeta'}
\end{align}
and by iterating  $\zeta''_{n}(\rho)$ can be estimated, finding
\begin{align}
\zeta''_{n}(\rho) & \simeq\left[\Lambda_{\star}\rho^{2}-2\sqrt{\Lambda_{\star}}\left(n+\frac{1}{2}\right)+\frac{n(n-1)}{\rho^{2}}\right]\zeta_{n}\left(\rho\right).\label{eq:zeta''}
\end{align}
The same procedure gives the following expressions for $\chi'_{k}\left(\rho,\,\phi\right)$ and $\chi''_{k}\left(\rho,\phi\right)$
\begin{gather}
\chi'_{k}\left(\rho,\,\phi\right)\simeq\left[-m_{\star}\phi^{2}\rho+\frac{k+\frac{1}{2}}{\rho}\right]\chi{}_{k}\left(\rho,\,\phi\right),\label{eq:chi'}
\end{gather}
\begin{align}
\chi''_{k}\left(\rho,\phi\right) & \simeq\left\{ m_{\star}^{2}\phi^{4}\rho^{2}-2m_{\star}\left(k+1\right)\phi^{2}+\frac{\left(k+\frac{1}{2}\right)\left(k-\frac{1}{2}\right)}{\rho^{2}}\right\} \chi{}_{k}\left(\rho,\,\phi\right).\label{eq: chi''}
\end{align}
By using the relations above, the eigenvalue problem \eqref{eq:WDW_chi'',zeta'',2chi'zeta'-1} becomes
\begin{equation}
\frac{3\pi \ell_P^4}{2 G}\zeta''_{n}\left(\rho\right)\chi_{k}\left(\rho,\,\phi\right)-\frac{\hbar}{2\rho^{2}}\zeta_{n}\left(\rho\right)\frac{\partial^{2}}{\partial\phi^{2}}\chi_{k}\left(\rho,\,\phi\right)+\Upsilon\left(\rho,\phi\right)\zeta_{n}\left(\rho\right)\chi_{k}\left(\rho,\,\phi\right)=E\,\zeta_{n}\left(\rho\right)\chi_{k}\left(\rho,\,\phi\right).\label{eq:eq_big_rho_final_bis}
\end{equation}
where
\begin{align}
\Upsilon\left(\rho,\phi\right)&=\underset{a}{\underbrace{\frac{\sqrt{3}}{2}\hbar\sqrt{\Lambda}\left(k+\frac{1}{2}\right)}}-\underset{b}{\underbrace{\frac{\Lambda}{8\pi G}\rho^{2}}}+\underset{c}{\underbrace{\frac{3\pi \ell_p^4}{2G}\frac{1}{\rho^{2}}\left(k+\frac{1}{2}\right)\left[2n+k-\frac{1}{2}\right]}}\nonumber\\
&+\underset{d}{\underbrace{\left(\frac{1}2\frac{m^{2}}{\hbar}+\frac{\sqrt{3}}{2}m\sqrt{\Lambda}
\right)\rho^{2}\phi^{2}}}-\underset{e}{\underbrace{3\pi\ell_P^2m\left(n+k+1\right)\phi^{2}}}+\underset{f}{\underbrace{\frac{3\pi\ell_P^2}{2\hbar}m^2\rho^{2}\phi^{4}}}.\label{eq:pot_big_rho}
\end{align}
We proceed to the analysis of each single term in \eqref{eq:pot_big_rho}:
\begin{description}

\item [{a)}] The constant terms redefine the energy eigenvalue;

\item [{b)}] The terms of $\rho^{2}$ order is the leading one as
$\rho\rightarrow\infty$;

\item [{c)}] The terms proportional to $\frac{1}{\rho^{2}}$ are negligible as soon as
\begin{equation}
\left|E-\frac{\sqrt{3}}{2}\hbar\sqrt{\Lambda}\left(k+\frac{1}{2}\right)\right|\gg\frac{3\pi\ell_P^4}{2G}\frac{1}{\rho^{2}}\left(k+\frac{1}{2}\right)\left(2n+k-\frac{1}{2}\right).\label{eq:1/rho^2 trasc}
\end{equation}

\item [{d)}] The terms $\rho^{2}\phi^{2}$ determine the quadratic part of the effective scalar field  potential and, to preserve the solution \eqref{eq:chi_k}, we must
impose  
\begin{equation}
m\gg\hbar\sqrt{3\Lambda},\label{eq:condiz_lambda}
\end{equation}
which tell us that the square root of the cosmological constant must be negligible with respect to the mass of the inflaton in the false vacuum.

\item [{e)}] The contribution of order $\rho^{0}\phi^{2}$ can be neglected with respect to
that of order $\rho^{2}\phi^{2}$ as soon as 
\begin{equation}
\frac12 m^2\rho^{2}\gg 3\pi\ell_P^2m\left(n+k+1\right).\label{eq:phi^2 trasc}
\end{equation}

\item [{f)}] The term proportional to $\rho^{2}\phi^{4}$ can be seen as a quartic potential contribution for the scalar field and it can be neglected if  
\begin{equation}
\phi^2\ll \frac{1}{3\pi\ell_P^2},
\end{equation}
which means that the scalar field takes sub-Planckian values (this is basically the reason why we cannot apply this framework to chaotic inflation \cite{Linde:1983gd}).
\end{description}

Therefore, as soon as $\rho$ is sufficiently big that the conditions \eqref{eq:1/rho^2 trasc} and \eqref{eq:phi^2 trasc} hold, Eq. \eqref{eq:WDW_chi'',zeta'',2chi'zeta'-1} reduces to 
\begin{align}
&\frac{3\pi \ell_P^4}{2 G}\zeta''_{n}\left(\rho\right)\chi_{k}\left(\rho,\,\phi\right)-\frac{\hbar}{2\rho^{2}}\zeta_{n}\left(\rho\right)\frac{\partial^{2}}{\partial\phi^{2}}\chi_{k}\left(\rho,\,\phi\right)+\nonumber\\&\left(-\frac{\Lambda}{8\pi G}\rho^{2}+\frac{1}2m^{2}\rho^{2}\phi^{2}+\frac{\sqrt{3}}{2}\hbar\sqrt{\Lambda}\left(k+\frac{1}{2}\right)\right)\zeta_{n}\left(\rho\right)\chi_{k}\left(\rho,\,\phi\right)=E\,\zeta_{n}\left(\rho\right)\chi_{k}\left(\rho,\,\phi\right).\label{eq:big_rho_true_final}
\end{align}
from which it follows that \eqref{eq:psi} is the proper eigenfunction with eigenvalue \eqref{eq:E_{n,k}-2}.
  
If the Universe energy eigenvalue is preserved, we can take for $E$ the value \eqref{maxspectpol} (for $\lambda_\star=\ell_P^{3/2}$) and estimate $n$ via the relation \eqref{eq:E_{n,k}-2} (we assume the contribution of the scalar field to be negligible). Hence, from \eqref{eq:1/rho^2 trasc} we get 
\begin{equation}
 \rho^2\gg 2n\,\ell_P^3\rightarrow a\gg 10^{-6}m,
\end{equation}
which fixes a lower bound for the age in which the decay into the true vacuum (thus also inflation) starts.

{\acknowledgments
The work of F. C. was supported by funds provided by the National Science Center under the agreement DEC12
2011/02/A/ST2/00294.}

\end{document}